\documentclass[aps,prd,superscriptaddress,twoside,twocolumn,nofootinbib,showpacs,floatfix]{revtex4-2}
\usepackage{amsmath,amssymb}
\usepackage{graphicx,bm,braket}
\usepackage{subfigure}
\usepackage{booktabs}
\usepackage{dcolumn}
\usepackage{multirow} 
\usepackage{array}
\usepackage{color}
\pdfpagewidth=\paperwidth
\pdfpageheight=\paperheight
\allowdisplaybreaks
\usepackage{epstopdf}

\newcommand*{\hc}{\text{H.\,c.}}

\usepackage{float}
\makeatletter

\newcommand{\Rmnum}[1]{\expandafter\@slowromancap\romannumeral #1@}
\makeatother
\usepackage{ulem}

\begin{document}

\title{\boldmath Decay behavior of hidden-strange hadronic molecular state $N(2270)$}

\author{Di Ben}
\email{bendi20@mails.ucas.ac.cn}
\affiliation{School of Nuclear Science and Technology, University of Chinese Academy of Sciences, Beijing 101408, China}
\affiliation{Institute of Theoretical Physics, Chinese Academy of Sciences, Beijing 100190, China}

\author{Shu-Ming Wu}
\email{wushuming@ucas.ac.cn}
\affiliation{School of Physical Sciences, University of Chinese Academy of Sciences, Beijing 100049, China}


\date{\today} 

\begin{abstract}
In this article, we systematically discuss the decay patterns of the hidden-strange hadronic molecular state $N(2270)$ which is assumed as an s-wave $K^\ast\Sigma^\ast$ shallow bound state with its possible quantum numbers $J^P$ which are ${1/2}^-$, ${3/2}^-$ and ${5/2}^-$. 
By using the effective Lagrangian approach and considering pseudo-scalar meson and vector meson exchanges, we have thoroughly calculated and discussed the decay behavior of the $N(2270)$ molecular state, including hadronic and radiative decay and the cutoff dependence, for different $J^P$ values.
For all three cases, the $K^\ast \Lambda$ final state is always included as the main decay channel.
However, the $K \Sigma$, $K \Lambda$, $K \Sigma^\ast$ and $\pi \Delta$ final states exhibit notable differences.
These different decay properties will provide valuable guidance for future experimental searches and aid in distinguishing different $J^P$ assumptions and understanding their internal structures.
\end{abstract}



\maketitle

\section{Introduction}\label{Sec:intro}

As the deeper understanding of baryons, multiquark states, including pentaquark states and hadronic molecular states, as well as hybrid states, are epicenters in hadron physics.
Different structures will lead to different natures, which are very interesting to study, and the internal structure is always an important route to get a deeper comprehension of the interactions of our physical world.

2015 was a breaking year, when the LHCb Collaboration presented striking evidence for $J/\psi \, p$ resonances, named $P_c^+(4380)$ and $P_c^+(4450)$, in $\Lambda^0_b\to K^- J/\psi \, p$ decays \cite{Aaij:2015tga}. 
Further information was reported in 2019, the LHCb Collaboration declared the $P_c^+(4312)$ state and a two-peak structure of the $P_c^+(4450)$ state, which is resolved into $P_c^+(4440)$ and $P_c^+(4457)$ \cite{Aaij:2019}. 

There are many theoretical investigations on the nature of the $P_c$ states. 
Before the experimental observation, many theoretical groups had predicted the existence of resonant structures near the $\bar{D}\Sigma_c$ and $\bar{D}^\ast\Sigma_c$ thresholds \cite{Wu:2010jy, Wu:2010vk, Wang:2011rga, Yang:2011wz, Wu:2012md, Yuan:2012wz, Xiao:2013yca, Uchino:2015uha, Karliner:2015ina}. When the pentaquark state was first discovered, many theoretical groups attempted to explain the structure of $P_c$ states by using the compact pentaquark state model while the different states are interpreted as $S$-wave and $P$-wave compact pentaquark states \cite{Chen:2016qju, Ali:2016dkf, Ali:2019npk, Maiani:2015vwa, Li:2015gta, Mironov:2015ica, Anisovich:2015cia, Zhu:2015bba, Ghosh:2015xqp, Wang:2015epa, Hiyama:2018ukv}.

The most intriguing aspect is that the reported masses of $P_c^+(4380)$ and $P_c^+(4457)$ are located just below the thresholds of $\bar{D}\Sigma_c^\ast$ and $\bar{D}^\ast\Sigma_c$ at $4382$ MeV and $4459$ MeV.
In the hadronic molecular picture, this strongly supports the interpretation of $P_c^+(4380)$ and $P_c^+(4457)$ as hadronic molecular states composed of $\bar{D} \Sigma_c^\ast$ and $\bar{D}^\ast \Sigma_c$, respectively~\cite{Guo:2017jvc, He:2019rva, Chen:2019asm, Liu:2019zvb, Du:2021fmf, Yalikun:2021bfm}. Furthermore, in a recent study~\cite{Wu:2024bvl}, the $P_c^+(4457)$ in the hadronic molecular picture can also be explained as the $\bar{D}^\ast \Lambda_c$ bound state.
Either $\bar{D} \Sigma_c^\ast$, $\bar{D}^\ast \Sigma_c$, or even $\bar{D}^\ast \Lambda_c$ system, which contains a pair of $c\Bar{c}$ quarks, is a so-called hidden charm in the hadronic molecular picture.

Analogously, in the light quark sector, the hidden-strange hadronic molecular system possibly exists too.
As the masses of $N(1875){3/2}^-$ and $N(2080){3/2}^-$ are just below the thresholds of $K\Sigma^\ast$ and $K^\ast\Sigma$ at $1880$ MeV and $2086$ MeV, respectively.
The $N(1875){3/2}^-$ and $N(2080){3/2}^-$ are proposed to be the strange partners of the $P_c^+(4380)$ and $P_c^+(4457)$ molecular states \cite{He:2017aps,Lin:2018kcc}. 
In Ref. \cite{Lin:2018kcc}, the decay patterns of $N(1875){3/2}^-$ and $N(2080){3/2}^-$ as s-wave $K\Sigma^\ast$ and $K^\ast\Sigma$ molecular states were calculated within an effective Lagrangian approach. It was found that the measured decay properties of $N(1875){3/2}^-$ and $N(2080){3/2}^-$ can be reproduced well, supporting the molecule interpretation of the $N(1875){3/2}^-$ and $N(2080){3/2}^-$ states.

Besides $N(1875){3/2}^-$ and $N(2080){3/2}^-$ states, in the hidden-strange molecular sector, some investigations of the state which is bounded by $K^\ast\Sigma^\ast$ named as $N(2270)$ are already presented in some photoproduction scattering processes.
In Ref. \cite{Ben:2023uev}, the hidden-strange molecular states $N(2080){3/2}^-$ and $N(2270){3/2}^-$ are compatible with the available cross-section data of the $\gamma p \rightarrow K^{\ast +}\Sigma^0$ and $\gamma p \rightarrow K^{\ast 0}\Sigma^+$ reactions.
What's more, in $\gamma p \rightarrow \phi p$ reactions \cite{Wu:2023ywu}, which also introduce $N(2080){3/2}^-$ and $N(2270){3/2}^-$ states in $s$-channel as its main mechanism, the experimental data are perfectly reproduced by the theoretical calculations.
In this way, the properties of $N(2270)$ should be an exhaustive study.

In this paper, we systematically discuss the decay patterns of the hidden-strange hadronic molecular state $N(2270)$ which is assumed as an s-wave $K^\ast\Sigma^\ast$ shallow bound state with possible quantum number $J^P$ of ${1/2}^-$, ${3/2}^-$ and ${5/2}^-$ by using the effective Lagrangian approach.

This paper is organized as follows. In Sec. \ref{Sec:forma}, we briefly introduce the framework of our theoretical model. In Sec. \ref{Sec:results}, the results of our theoretical calculations with some discussions are presented. Finally, we give a brief summary and conclusions in Sec. \ref{sec:summary}.

\section{Formalism}\label{Sec:forma}

\begin{figure}[htb]
    \centering
    \includegraphics[width=0.27\textwidth]{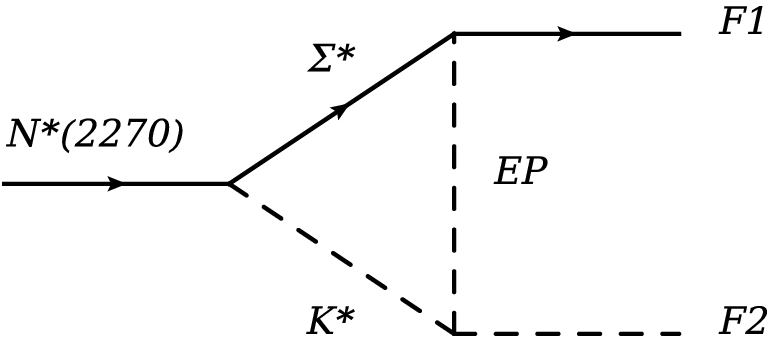}
    \vglue 6pt
    \caption{The triangle diagrams of $N(2270)$ decays, where F1 and F2 denote the final states and EP denotes the exchange particles.}
    \label{2270}
\end{figure}

Following the calculations in Ref. \cite{Lin:2018kcc},
the decay of the $N(2270)$ molecular state proceeds through triangular diagrams, as shown in Fig. \ref{2270}, where F1 and F2 denote the final states and EP denotes the exchange particles.
Considering the exchange of pseudo-scalar and vector mesons, we list the possible hadronic decay channels and corresponding exchange particles in Tab. \ref{tab:decaychannel}.

\begin{table}[htb]
    \caption{Hadronic Decay Modes of $N^*(2270)(K^*\Sigma^*)$}
    \belowrulesep=0pt
\aboverulesep=0pt
\renewcommand{\arraystretch}{1.5}
\centering
\begin{tabular}{c|c}
\toprule[1pt]
    \makebox[0.1\textwidth][c]{Final states} & \makebox[0.2\textwidth][c]{Exchanged particles}\\\hline
    $K\Sigma$ & $\rho,\omega,\pi$ \\
    $K\Lambda$ & $\rho,\pi$ \\
    $K\Sigma^*$ & $\rho,\omega,\pi$ \\
    $K^*\Sigma$ & $\rho,\omega,\pi$ \\
    $K^*\Lambda$ & $\rho,\pi$ \\
    $\pi N$ & $K,K^*$\\
    $\pi \Delta$ & $K,K^*$\\
    $\eta N$ & $K,K^*$\\
    $\rho N$ & $K,K^*$\\
    $\omega N$ & $K,K^*$\\
    $\phi N$ & $K,K^*$\\
    \bottomrule[1pt]
\end{tabular}
    \label{tab:decaychannel}
\end{table}

\subsection{Lagrangians}

$N(2270)$ as an s-wave $K^* \Sigma^*$ molecular state with the assuming mass $m_{N(2270)}=2270$ MeV.  
The $J^P$ of $K^*$ is $1^-$, while the $J^P$ of $\Sigma^*$ is ${3/2}^+$. For the s-wave coupling to the $N(2270)$ states, the allowed $J^P$s of $N(2270)$ are ${1/2}^-$, ${3/2}^-$ and ${5/2}^-$.
By the Lorentz covariant orbital-spin scheme~\cite{Zou:2002yy}, the s-wave couplings of $N(2270)$ with $K^* \Sigma^*$ for different $J^P$  are given by:

\begin{equation}
    \mathcal{L}_{K^* \Sigma^* N^*({1/2}^-)} = g^{{1/2}^-}{\Sigma^{*}}^\mu N^* {K^*}_\mu,
    \label{1/2}
\end{equation}
\begin{equation}
    \mathcal{L}_{K^* \Sigma^* N^*({3/2}^-)} = g^{{3/2}^-}{\Sigma^{*}}^\mu \gamma^5\Tilde{\gamma}^\nu {N^*}_\mu {K^*}_\nu,
    \label{3/2}
\end{equation}
\begin{equation}
    \mathcal{L}_{K^* \Sigma^* N^*({5/2}^-)} = g^{{5/2}^-}{\Sigma^{*}}^\mu {N^*}_{\mu\nu} {K^*}^\nu,
\end{equation}
where $g^{{1/2}^-}$, $g^{{3/2}^-}$ and $g^{{5/2}^-}$ corresponding to the coupling constants when $\Sigma^*$ and $K^*$ field coupled to the different spin parity $N^*$ field with $J^P$ as ${1/2}^-$, ${3/2}^-$ and ${5/2}^-$ respectively.

In the discussion below, to simplify, $V$ stands for vector meson fields and $P$ stands for pseudo-scalar meson fields, while $D$ stands for the baryon-decuplet fields and $B$ stands for the baryon-octet fields. Besides the hadronic molecular vertex, the Lagrangians are given~\cite{Oset:2010tof, Ronchen:2012eg, Wu:2010vk, Matsuyama:2006rp}:

\begin{equation}
    \mathcal{L}_{VPP} = ig_{VP_1P_2}(V_\mu\partial^\mu P_1 P_2-V_\mu\partial^\mu P_2 P_1),
\end{equation}
\begin{equation}
\begin{aligned}
    \mathcal{L}_{VVV} = &-ig_{V_1V_2V_3}({V_1}^\mu(\partial_\mu {V_2}^\nu{V_3}_\nu-{V_2}^\nu\partial_\mu{V_3}_\nu)\\
    &+(\partial_\mu {V_1}^\nu{V_2}_\nu-{V_1}^\nu\partial_\mu{V_2}_\nu){V_3}^\mu)\\
    &+{V_2}^\mu({V_1}^\nu\partial_\mu{V_3}_\nu-\partial_\mu {V_1}^\nu{V_3}_\nu)),
\end{aligned}
\end{equation}
\begin{equation}
    \mathcal{L}_{VVP} = -g_{V_1V_2P}\epsilon^{\mu\nu\alpha\beta}(\partial_\mu {V_1}_\nu\partial_\alpha {V_2}_\beta)P,
\end{equation}
\begin{equation}
        \mathcal{L}_{DBP} = g_{DBP}\Bar{D}^\mu \partial_\mu P B + \hc,
\end{equation}
\begin{equation}
    \mathcal{L}_{DBV} = -ig_{DBV}\Bar{D}^\mu\gamma^\nu\gamma_5(\partial_\mu V_\nu-\partial_\nu V_\mu)B + \hc,
\end{equation}
\begin{equation}
    \mathcal{L}_{DDP} = \frac{f_{DDP}}{m_\pi}\Bar{D_1}^\mu\gamma^\nu {D_2}_\mu\partial_\nu P,
\end{equation}
\begin{equation}
    \mathcal{L}_{DDV} = g_{DDV}\Bar{D_1}^\alpha(\gamma^\mu-\frac{\kappa_{DDV}}{2m_\Delta}\sigma^{\mu\nu}\partial_\nu)V_\mu{D_2}_\alpha,
\end{equation}
where $g_{VP_1P_2}$, $g_{V_1V_2V_3}$ and $g_{V_1V_2P}$ are coupling constants, $g_{DBP}$, $g_{DBV}$, $f_{DDP}$, $g_{DDV}$ and $\kappa_{DDV}$ are coupling constants. $m_\pi$ and $m_\Delta$ are the pion and $\Delta$ mass.

\subsection{Coupling Constants}\label{Cons}

Following Ref. \cite{Lin:2019qiv}, we can calculate the effective s-wave coupling constants with the Weinberg compositeness criterion:

\begin{equation}
    g_0(J^P) = \sqrt{\frac{8\sqrt{2}\sqrt{E_B}m_1m_2\pi}{(m_1m_2/(m_1+m_2))^{3/2}}}\sqrt{\frac{1}{M_N(J^P)F_T(J^P)}}.
    \label{1}
\end{equation}

\begin{table}[htb]
    \caption{$M_N$ and $F_T$ values}
\belowrulesep=0pt
    \aboverulesep=0pt
    \renewcommand{\arraystretch}{1.5}
    \centering
\begin{tabular}{c|c|c|c}
    \toprule[1pt]
    \makebox[0.1\textwidth][c]{$J^P$} & \makebox[0.1\textwidth][c]{$M_N$}& \makebox[0.1\textwidth][c]{$F_T$} &\makebox[0.1\textwidth][c]{$g_0$}\\\hline
    ${1/2}^-$ & $4m_1$& $1$ &$1.340$\\
    ${3/2}^-$ & $20/9m_1$& $3/2$ &$1.468$\\
    ${5/2}^-$ & $6/5m_1$& $5/3$ &$1.895$\\
    \bottomrule[1pt]
\end{tabular}
    \label{tab:BE}
\end{table}

For different quantum numbers $J^P$, $M_N$, and $F_T$ have different values. The calculated $g_0(J^P)$ are listed in the Tab. \ref{tab:BE}, where binding energy $E_B = 8.19$ MeV, $m_2 = m_{K^*} = 893.61$ MeV and $m_1 = m_{\Sigma^*} = 1384.58$ MeV.

All other coupling constants can be calculated by the $SU(3)$ relations and are listed in Tab. \ref{tbl:table1}. 
It is important to note that due to different phase conventions, the $SU(3)$ relations may vary across different articles, as discussed in detail in Ref. \cite{Lu:2024ajt}.
Therefore, we adopt the conventions used by de Swart~\cite{deSwart:1963pdg, ParticleDataGroup:2024cfk}.

\begin{table*}
\belowrulesep=0pt
\aboverulesep=0pt
\renewcommand{\arraystretch}{2.5}
\centering
\caption{Coupling Constants}
\begin{tabular}{ccc|ccc}
    \toprule[1pt]
    \makebox[0.1\textwidth][c]{Coupling} 
    & \makebox[0.2\textwidth][c]{ $SU(3)$ relations}
    & \makebox[0.1\textwidth][c]{Values} 
    & \makebox[0.1\textwidth][c]{Coupling}
    & \makebox[0.2\textwidth][c]{ $SU(3)$ relations}
    & \makebox[0.1\textwidth][c]{Values} \\
\midrule
\hline 
 $ g_{\rho \rightarrow \omega \pi }$ & $ 2\sqrt{3} g_{VVP}$ & $ 12.8\mathrm{~GeV}^{-1}$~\cite{Nakayama:2006ps} & $ g_{\rho \rightarrow \pi \pi }$ & $ 2\sqrt{2} g_{VPP}$ & $ 8.54$~\cite{Janssen:1994wn} \\
$ g_{K^{*}\rightarrow \rho K }$ & $ -3 g_{VVP}$ & $ -11.1\mathrm{~GeV}^{-1}$ & $ g_{K^{*}\rightarrow K\pi }$ & $ \sqrt{3} g_{VPP}$ & $ 5.23$ \\
$ g_{K^{*}\rightarrow \omega K}$ & $ \sqrt{3} g_{VVP}$ & $ 6.42\mathrm{~GeV}^{-1}$ & $ g_{K^{*}\rightarrow K\eta }$ & $ \sqrt{3} g_{VPP}$ & $ 5.23$ \\
\cline{4-6} 
 $ g_{K^{*}\rightarrow \phi K}$ & $ \sqrt{6} g_{VVP}$ & $ 9.08\mathrm{~GeV}^{-1}$ & $ g_{\Delta \rightarrow \Delta \rho }$ & $ \sqrt{15} g_{DDV}$ & $ 9.91$~\cite{Schutz:1998jx} \\
$ g_{K^{*}\rightarrow K^{*} \pi }$ & $ 3g_{VVP}$ & $ 11.1\mathrm{~GeV}^{-1}$ & $ g_{\Sigma ^{*}\rightarrow \Sigma ^{*} \rho }$ & $ 2\sqrt{2} g_{DDV}$ & $ 7.24$ \\
$ g_{K^{*}\rightarrow K^{*} \eta }$ & $ -g_{VVP}$ & $ -3.71\mathrm{~GeV}^{-1}$ & $ g_{\Delta \rightarrow \Sigma ^{*} K^{*}}$ & $ -\sqrt{6} g_{DDV}$ & -$ 6.27$ \\
\cline{1-3} 
 $ g_{\rho \rightarrow \rho \rho }$ & $ 2\sqrt{2} g_{VVV}$ & $ 8.28$~\cite{Oset:2010tof} & $ g_{\Sigma ^{*}\rightarrow \Sigma ^{*} \omega }$ & $ 2g_{DDV}$ & $ 5.12$ \\
$ g_{K^{*}\rightarrow K^{*} \rho }$ & $ \sqrt{3} g_{VVV}$ & $ 5.07$ &  & $ \kappa _{DDV}$ & $ 6.1$~\cite{Schutz:1998jx} \\
\cline{4-6} 
 $ g_{K^{*}\rightarrow K^{*} \omega }$ & $ g_{VVV}$ & $ 2.93$ & $ f_{\Delta \rightarrow \Delta \pi }$ & $ \sqrt{15} f_{DDP}$ & $ 2.30$~\cite{Schutz:1998jx} \\
$ g_{K^{*}\rightarrow K^{*} \phi }$ & $ -\sqrt{2} g_{VVV}$ & $ -4.14$ & $ f_{\Sigma ^{*}\rightarrow \Sigma ^{*} \pi }$ & $ 2\sqrt{2} f_{DDP}$ & $ 1.68$ \\
\cline{1-3} 
 $ g_{N\rightarrow \Delta \rho }$ & $ -2\sqrt{3} g_{BDV}$ & $ 29.24\mathrm{~GeV}^{-1}$~\cite{Janssen:1996kx} & $ f_{\Delta \rightarrow \Sigma ^{*} K}$ & $ -\sqrt{6} f_{DDP}$ & -$ 1.45$ \\
\cline{4-6} 
 $ g_{N\rightarrow \Sigma ^{*} K}$ & $ \sqrt{3} g_{BDV}$ & $ -14.26\mathrm{~GeV}^{-1}$ & $ g_{N\rightarrow \Delta \pi }$ & $ -2\sqrt{3} g_{BDP}$ & $ 21.79\mathrm{~GeV}^{-1}$~\cite{Janssen:1996kx} \\
$ g_{\Sigma \rightarrow \Sigma ^{*} \rho }$ & $ -\sqrt{2} g_{BDV}$ & $ 11.94\mathrm{~GeV}^{-1}$ & $ g_{N\rightarrow \Sigma ^{*} K}$ & $ \sqrt{3} g_{BDP}$ & $ -10.90\mathrm{~GeV}^{-1}$ \\
$ g_{\Lambda \rightarrow \Sigma ^{*} \rho }$ & $ -3g_{BDV}$ & $ 25.33\mathrm{~GeV}^{-1}$ & $ g_{\Sigma \rightarrow \Sigma ^{*} \pi }$ & $ $ & $ 6.030\mathrm{~GeV}^{-1}$ \\
$ g_{\Sigma \rightarrow \Sigma ^{*} \omega }$ & $ -g_{BDV}$ & $ 8.442\mathrm{~GeV}^{-1}$ & $ g_{\Lambda \rightarrow \Sigma ^{*} \pi }$ & $ $$ $ & $ 13.39\mathrm{~GeV}^{-1}$ \\
\bottomrule[1pt]
\end{tabular}
\label{tbl:table1}
\end{table*} 

Additionally, for the couplings $g_{\Sigma \rightarrow \Sigma ^{*} \pi }$ and $g_{\Lambda \rightarrow \Sigma ^{*} \pi }$, we did not use the values provided by the $SU(3)$ relations. Instead, we used values calculated from the widths listed in the PDG~\cite{ParticleDataGroup:2024cfk}, ensuring that the signs are consistent with those in $SU(3)$ relations.

For $g_{\Sigma^\ast\to\Sigma^\ast \omega}$ and $g_{\Sigma\to\Sigma^\ast \omega}$, we discussed separately.
Assuming ideal mixing of $\omega_1$ and $\omega_8$,
\begin{align}
    \omega &= \sqrt{\frac{2}{3}}\omega_1+\sqrt{\frac{1}{3}}\omega_8, \\
    \phi &= \sqrt{\frac{1}{3}}\omega_1-\sqrt{\frac{2}{3}}\omega_8.
\end{align}

According to $SU(3)$ relations, we will find:
\begin{gather}
    g_{\Sigma^\ast \to \Sigma^\ast \omega_8} = 0,\\
    g_{\Delta\to\Delta\omega_8} = \sqrt{\frac{3}{15}} g_{\Delta\to\Delta\rho}, \\
    g_{\Sigma^\ast \to\Sigma^\ast \omega_1} = g_{\Delta\to\Delta\omega_1},\\
    g_{\Sigma\to\Sigma^\ast \omega_8} = \frac{1}{2}g_{N\to\Delta\rho},\\
    g_{\Sigma\to\Sigma^\ast \omega_1} = 0.
\end{gather}

Furthermore, we assume that the $\phi$ meson does not couple to $\Delta$ (OZI rule), i.e., $g_{\Delta\to\Delta\phi} = 0$. Then, we have
\begin{gather}
    g_{\Delta\to\Delta\omega_1} = \sqrt{2}g_{\Delta\to\Delta\omega_8} = \sqrt{\frac{2}{5}} g_{\Delta\to\Delta\rho},
\end{gather}

So,
\begin{align}
    g_{\Sigma^\ast\to\Sigma^\ast \omega} &= \sqrt{\frac{2}{3}}g_{\Sigma^\ast\to\Sigma^\ast \omega_1} = \frac{2}{\sqrt{15}} g_{\Delta\to\Delta\rho}, \\
    g_{\Sigma\to\Sigma^\ast \omega} &= \sqrt{\frac{1}{3}}g_{\Sigma\to\Sigma^\ast \omega_8} = \frac{1}{2\sqrt{3}} g_{N\to\Delta\rho}.
\end{align}













The tensor term should also be determined in Ref. \cite{Meng:2017fwb, Nakayama:2006ps, Schutz:1998jx}. 
According to the $SU(3)$ isospin symmetry, the tensor coupling should also be changed by the isospin factor just like the $g_{DDV}$ does, so the value of $\kappa_{DDV}$ should always be the same under the $SU(3)$ isospin symmetry, thus this value is taken as $\kappa_{DDV}=6.1$.


\subsection{Form Factors}\label{ff}

To avoid ultraviolet (UV) divergence and to suppress short-distance contributions in the calculation, we introduce the following Gaussian regulator to render all amplitudes UV finite:

\begin{equation}
    f(\textbf{q}^2/{\Lambda_0}^2) = \text{exp}(-\textbf{q}^2/{\Lambda_0}^2).
\end{equation}

Where $\textbf{q}$ is the spatial part of loop momentum $l$, and $\Lambda_0$ is a UV cutoff.

In addition, a multi-pole regulator is introduced to suppress the off-shell contribution for the exchange particles, which is:

\begin{equation}
    f(l^2) = \frac{\Lambda^4}{(m^2-l^2)^2+\Lambda^4}.
\end{equation}

Where $m$ is the mass of the exchanged particle and $l$ is the corresponding four-momentum, while $\Lambda$ is the cutoff parameter.

Two cutoff parameters, one of which is the $\Lambda_0$ in the Gaussian regulator to normalize UV divergence, and the other one is the $\Lambda$ introduced by the multi-poles regulator to suppress the off-shell effect.
In general, the values for these two cutoffs are taken around the value which is much larger than the $\Lambda_{\text{QCD}}$, so we take $\Lambda_0$ and $\Lambda$ as $900$ MeV.
When it comes to the discussion of model dependence of cutoffs, we will set one of the cutoffs as $900$ MeV and shift the other one from $800$ MeV to $1000$ MeV.

\subsection{Radiative Decay}\label{PR}
To calculate the radiative decay, we consider the vector mesons dominant(VMD) model with the following Feynman diagram shown in Fig. \ref{2270photon}:

\begin{figure}[htb]
    \centering
    \includegraphics[width=0.32\textwidth]{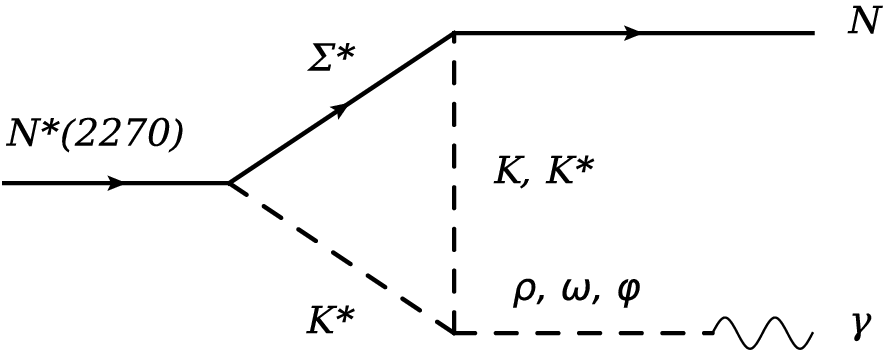}
    \vglue 6pt
    \caption{The triangle diagrams of $N(2270)$ radiative decay.}
    \label{2270photon}
\end{figure}

The VMD is defined by the following Lagrangian~\cite{Wu:2019adv}:

\begin{equation}
    \mathcal{L}_{\text{VMD}}(x)=\frac{e{m_V^2}}{f_V}A_\mu(x)\phi^\mu_V(x).
\end{equation}

Where $m_V$ is the mass of the vector meson $V$, and $A_\mu$ and $\phi^\mu_V$ are the field operators for the photon and vector meson, respectively. The width of $V\rightarrow e^+e^-$ can then be calculated by:

\begin{equation}
    \Gamma_{V\rightarrow e^+e^-} = \frac{1}{3}\alpha^2 m_V \frac{4\pi}{{f_V}^2}.
\end{equation}

Where $\alpha$ is the fine structure constant. By using the data of $\Gamma_{V\rightarrow e^+e^-}$~\cite{ParticleDataGroup:2024cfk}, the decay constants can be determined: $f_\rho=5.33$, $f_\omega=15.2$, $f_\phi=13.4$.





\section{Results and Discussion}\label{Sec:results}

\subsection{Hadronic Decays}

As shown in Tab. \ref{tab:decay}, the decay behavior of the $N(2270)$ molecular state under the set of cutoffs $(\Lambda, \Lambda_0)$ in the calculation is $(900,900)$ MeV. 
Where the first column FS represents the final states of hadronic decays, the subscript of $\Gamma$ indicates the exchanged pseudo-scalar or vector meson, the last but one column $\Gamma$ represents the total decay width of this final state, and the last column represents the branching ratio ($\text{BR} = \Gamma_i/\Gamma$) of this final state.

The main decay channel is concluded as which BR is near or much larger than $10\%$.

\begin{table*}
    \caption{Decay width of $N^*(2270)(K^*\Sigma^*)$ with $J^P = 1/2^-$, $3/2^-$, and $5/2^-$ when $(\Lambda, \Lambda_0) = (900,900)$ MeV. Where the first column $FS$ represents the final states of hadronic decays, the subscript of $\Gamma$ indicates the exchanged pseudo-scalar or vector meson, the last but one column $\Gamma$ represents the total decay width of this final state, and the last column $\Gamma_i/\Gamma$ represents the branching ratio of this final state.}
    \belowrulesep=0pt
    \aboverulesep=0pt
    \renewcommand{\arraystretch}{1.8}
    \centering
    \begin{tabular}{c||c|c|c||c|c|c||c|c|c||c|c|c||c|c|c}
\toprule[1pt]
        FS & \multicolumn{3}{c||}{$\Gamma_\pi$(MeV)} & \multicolumn{3}{c||}{$\Gamma_\rho$(MeV)} & \multicolumn{3}{c||}{$\Gamma_\omega$(MeV)} & \multicolumn{3}{c||}{$\Gamma$(MeV)} & \multicolumn{3}{c}{$\Gamma_i/\Gamma$(\%)}\\\hline
         $(J^P)$ &~ $1/2^-$ &~ $3/2^-$ &~ $5/2^-$ &~ $1/2^-$ &~ $3/2^-$ &~ $5/2^-$ &~ $1/2^-$ &~ $3/2^-$ &~ $5/2^-$ &~ $1/2^-$ &~ $3/2^-$ &~ $5/2^-$ &~ $1/2^-$ &~ $3/2^-$ &~ $5/2^-$ \\
        \midrule[1pt]
        $K\Sigma$ 
        & $0.91$ & $0.69$ & $3.05$ 
        & $10.22$ & $2.15$ & $0.81$ 
        & $2.58$ & $0.66$ & $0.20$  
        & $29.35$ & $6.67$ & $2.02$ 
        & $11.85$ & $1.82$ & $2.56$ 
        \\[3pt]
        $K\Lambda$ 
        & $3.24$ & $4.56$ & $20.47$ 
        & $73.21$ & $17.43$ & $5.76$ 
        & --- & --- & --- 
        & $93.60$ & $25.15$ & $13.37$ 
        & $37.80$ & $6.88$ & $16.95$
        \\[3pt]
        $K\Sigma^*$ 
        & $2.93$ & $5.75$ & $6.15$ 
        & $2.87$ & $18.04$ & $3.20$ 
        & $0.89$ & $4.30$ & $0.83$  
        & $9.90$ & $45.67$ & $13.42$ 
        & $4.00$ & $12.50$ & $17.02$ 
        \\[3pt]
        $K^*\Sigma$ 
        & $3.27$ & $7.57$ & $2.73$ 
        & $5.50$ & $7.36$ & $2.36$ 
        & $2.11$ & $1.81$ & $0.57$  
        & $2.13$ & $14.32$ & $1.74$ 
        & $0.86$ & $3.91$ & $2.21$ 
        \\[3pt]
        $K^*\Lambda$ 
        & $28.00$ & $61.38$ & $22.72$ 
        & $46.63$ & $59.24$ & $21.00$ 
        & ---& --- & --- 
        & $24.68$ & $194.66$ & $18.94$ 
        & $9.97$ & $53.26$ & $24.01$ 
        \\
        \midrule[1pt]
        FS & \multicolumn{3}{c||}{$\Gamma_K$(MeV)} & \multicolumn{6}{c||}{$\Gamma_{K^*}$(MeV)} & \multicolumn{3}{c||}{$\Gamma$(MeV)} & \multicolumn{3}{c}{$\Gamma_i/\Gamma$(\%)}\\\hline
         $(J^P)$ &~ $1/2^-$ &~ $3/2^-$ &~ $5/2^-$ & \multicolumn{2}{c|}{$1/2^-$} & \multicolumn{2}{c|}{$3/2^-$} & \multicolumn{2}{c||}{$5/2^-$} &~ $1/2^-$ &~ $3/2^-$ &~ $5/2^-$ &~ $1/2^-$ &~ $3/2^-$ &~ $5/2^-$ \\
         \midrule[1pt]
        $\pi N$ 
        & $4.21$ & $0.41$ & $1.04$ 
        & \multicolumn{2}{c|}{$6.29$} & \multicolumn{2}{c|}{$2.69$} & \multicolumn{2}{c||}{$0.55$} 
        & $16.98$ & $2.35$ & $1.55$ 
        & $6.86$ & $0.64$ & $1.97$ 
        \\[3pt]
        $\pi \Delta$ 
        & $2.14$ & $5.19$ & $4.82$ 
        & \multicolumn{2}{c|}{$8.78$} & \multicolumn{2}{c|}{$35.14$} & \multicolumn{2}{c||}{$7.86$} 
        & $10.93$ & $40.32$ & $12.69$ 
        & $4.41$ & $11.03$ & $16.08$ 
        \\[3pt]
        $\eta N$ 
        & $4.72$ & $1.00$ & $3.22$ 
        & \multicolumn{2}{c|}{$1.49$} & \multicolumn{2}{c|}{$0.64$} & \multicolumn{2}{c||}{$0.13$} 
        & $2.51$ & $2.24$ & $3.10$ 
        & $1.01$ & $0.61$ & $3.93$ 
        \\[3pt]
        $\rho N$
        & $2.88$ & $3.15$ & $1.06$ 
        & \multicolumn{2}{c|}{$12.35$} & \multicolumn{2}{c|}{$4.69$} & \multicolumn{2}{c||}{$1.52$} 
        & $23.06$ & $11.08$ & $3.46$ 
        & $9.31$ & $3.03$ & $4.38$ 
        \\[3pt]
        $\omega N$ 
        & $2.85$ & $3.17$ & $1.06$ 
        & \multicolumn{2}{c|}{$12.17$} & \multicolumn{2}{c|}{$4.60$} & \multicolumn{2}{c||}{$1.51$} 
        & $7.25$ & $4.51$ & $1.68$ 
        & $2.93$ & $1.23$ & $2.13$ 
        \\[3pt]
        $\phi N$ 
        & $3.82$ & $6.78$ & $2.36$ 
        & \multicolumn{2}{c|}{$12.85$} & \multicolumn{2}{c|}{$5.12$} & \multicolumn{2}{c||}{$2.53$} 
        & $26.94$ & $18.49$ & $6.90$ 
        & $10.88$ & $5.06$ & $8.74$ \\
        \bottomrule[1pt]
    \end{tabular}
    \label{tab:decay}
\end{table*}

The result of molecular state $N(2270)$ with spin parity $J^P=1/2^-$ can be selected from Tab. \ref{tab:decay}. 
It shows that the most important decay channel of $N(2270)1/2^-$ is $K\Lambda$, with a BR $\sim 40\%$. Meanwhile,
other main decay channels include $K\Sigma$, $K^*\Lambda$, $\rho N$, and $\phi N$, and the decay BRs of these decay channels are about $10\%$.

The main decay channels are alternated for the $N(2270)$ with $J^P=3/2^-$. 
The $K^*\Lambda$ final state shows the absolute domination in this pattern, nearly half of the BR contribution from this channel. What's interesting is that $K\Sigma^\ast$ and $\pi\Delta$ are in the following position with the decay BR $\sim 10\%$. 

Finally, for the $N(2270)$ with $J^P=5/2^-$, the main decay channels are $K^*\Lambda$, $K\Sigma^*$, $K\Lambda$, $\pi\Delta$ and $\phi N$, and the results obtained for the other decay channels are relatively small. 
Also, the $K^*\Lambda$ final state shows the absolute domination, nearly $25\%$ of the BR contribution from this channel. Meanwhile, $K\Sigma^*$, $\pi\Delta$, and $K\Lambda$, contribute BR $\sim 17\%$ of $N(2270)5/2^-$ decays.

\subsection{Radiative Decay}

\begin{table}[H]
    \caption{The total width, radiative decay width, and branching ratio of $N(2270)$ when $(\Lambda, \Lambda_0) = (900,900)$ MeV. }
    \belowrulesep=0pt
    \aboverulesep=0pt
    \renewcommand{\arraystretch}{1.5}
    \centering
\begin{tabular}{c|c|c|c}
\toprule[1pt]
    \makebox[0.15\textwidth][c]{Initial state} & \makebox[0.1\textwidth][c]{$\Gamma$(MeV)} & \makebox[0.1\textwidth][c]{$\Gamma_{\gamma N}$(KeV) }& \makebox[0.1\textwidth][c]{$\Gamma_{\gamma N}/\Gamma$(\%)}\\\hline
    $N^*(2270)1/2^-$ & $247.61$ & $287.17$ & $0.116$\\
    $N^*(2270)3/2^-$ & $365.52$& $59.60$ & $0.016$\\
    $N^*(2270)5/2^-$ & $78.89$& $18.42$ & $0.023$\\\bottomrule[1pt]
    \end{tabular}
    \label{tab:totalwidth}
\end{table}

With the two cutoff parameters manually set to $(\Lambda, \Lambda_0) = (900,900)$ MeV, the calculated results are shown in Tab. \ref{tab:totalwidth}, the first column is the quantum numbers of all possible s-wave molecular states $N(2270)$ that we analyzed from the spin parity of $K^*$ and $\Sigma^*$ system, which are $1/2^-$, $3/2^-$ and $5/2^-$, respectively.
The second column shows the total width we calculated.
The third column shows the calculated width from radiative decay.
In this part, we use the VMD model and consider the exchange of pseudo-scalar and vector mesons in the triangle loop. 
Refer to Sec. \ref{PR}, the radiative decay width is the sum of the six Feynman diagrams listed. 
The last column gives the branching ratio of radiative decay.

As shown in Tab. \ref{tab:totalwidth}, when the initial spin parity $J^P = 1/2^-$, the BR of radiative decay is about $0.1\%$, and when the initial spin parity $J^P = 3/2^-$ or $5/2^-$, the BR of radiative decay is about $0.02\%$.

\subsection{Cutoff Dependence}

There are two cutoff parameters introduced in Sec. \ref{ff}.
Below, we will discuss the dependence of the results on the cutoff parameters one by one.

\subsubsection{Total Width}

\begin{figure}[H]
    \centering
    {\vglue 0.15cm}
    \subfigure{
    \includegraphics[width=0.77\columnwidth]{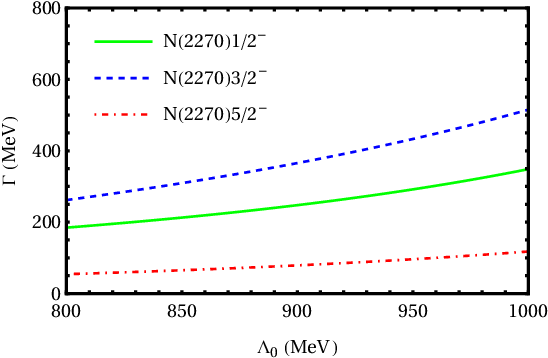}}
    \subfigure{
    \includegraphics[width=0.77\columnwidth]{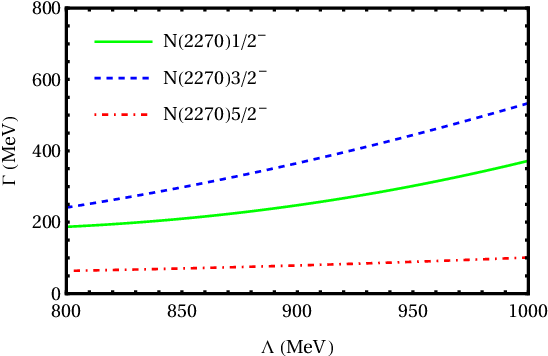}}
    \caption{Total width for $N(2270)$ as a function of cutoff, with $\Lambda=900$ MeV (upper figure) and $\Lambda_0=900$ MeV (lower figure). The green line, blue dashed line, and red dot-dashed line correspond to the case of initial state $N(2270)$ spin-parity quantum number $J^P$ as $1/2^-$, $3/2^-$, and $5/2^-$, respectively.}
    \label{fig:totalwidth}
\end{figure}

As shown in Fig. \ref{fig:totalwidth}, the upper figure is the result of fixing the $\Lambda$ as $900$ MeV and changing $\Lambda_0$ from $800$ MeV to $1000$ MeV, while the lower figure is the result of the fixing the $\Lambda_0$ as $900$ MeV and changing $\Lambda$ from $800$ MeV to $1000$ MeV. 
Where the green line, blue dashed line, and red dot-dashed line correspond to the case of initial state $N(2270)$ spin-parity quantum number $J^P$ as $1/2^-$, $3/2^-$, and $5/2^-$, respectively.

As Fig. \ref{fig:totalwidth} shows, with the vary of the cutoff parameter, the dependence of the total width on the cutoff parameter is very obvious, and the width can vary from $200$ to $500$ MeV, which can cover the possible width range of the general nucleon excited states.
Therefore, it is difficult for us to accurately predict the total width from the existing model, since the shifting range is large.

\subsubsection{Hadronic Decays}

\begin{figure*}[htb]
    \centering
    {\vglue 0.15cm}
    \subfigure{
    \includegraphics[width=0.855\columnwidth]{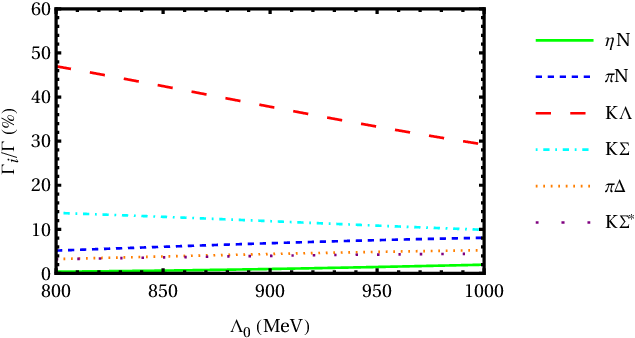}} {\hglue 0.4cm}
    \subfigure{
    \includegraphics[width=0.855\columnwidth]{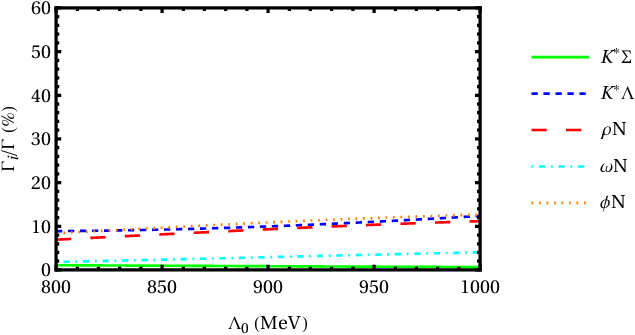}}\\
    \subfigure{
    \includegraphics[width=0.861\columnwidth]{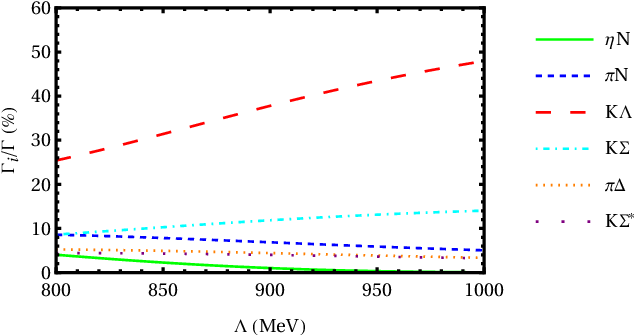}} {\hglue 0.32cm}
    \subfigure{
    \includegraphics[width=0.861\columnwidth]{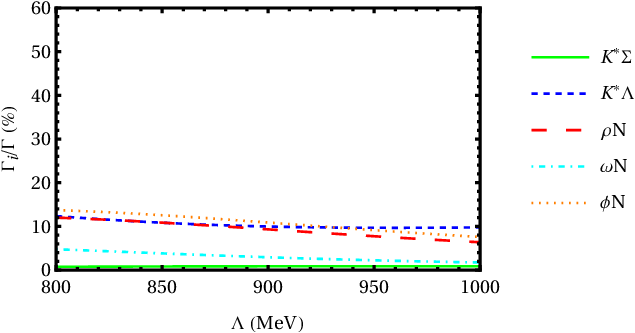}}
    \caption{Branching ratio for the $N(2270)1/2^-$ hadron decay as a function of cutoff, with $\Lambda = 900$ MeV (left panel) and $\Lambda_0 = 900$ MeV (right panel). We draw the branching ratios of the decay channels of the final states containing pseudo-scalar mesons in the first row, where the green lines, blue dashed lines, red large-dashed lines, cyan dot-dashed lines, orange dotted lines, and purple large-dotted lines represent $\eta N$, $\pi N$, $K\Lambda$, $K\Sigma$, $\pi \Delta$ and $K\Sigma^\ast$ final state respectively, and all the cases of involving vector mesons in the second row, where the green lines, blue dashed lines, red large-dashed lines, cyan dot-dashed lines and orange dotted lines represent $K^\ast \Sigma$, $K^\ast \Lambda$, $\rho N$, $\omega N$ and $\phi N$ final state respectively.}
    \label{fig:1h}
\end{figure*}

\begin{figure*}[htb]
    \centering
    {\vglue 0.15cm}
    \subfigure{
    \includegraphics[width=0.855\columnwidth]{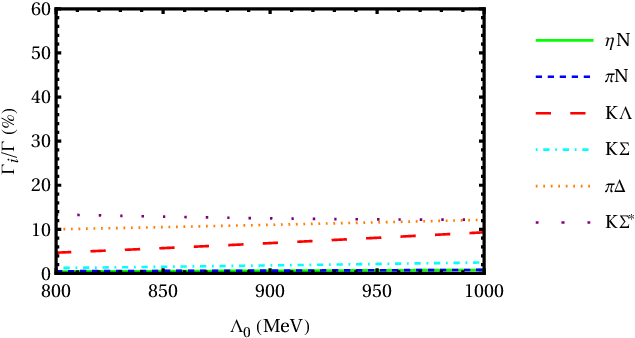}} {\hglue 0.4cm}
    \subfigure{
    \includegraphics[width=0.855\columnwidth]{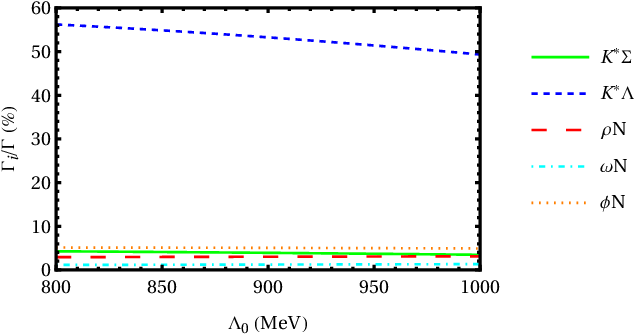}}\\
    \subfigure{
    \includegraphics[width=0.861\columnwidth]{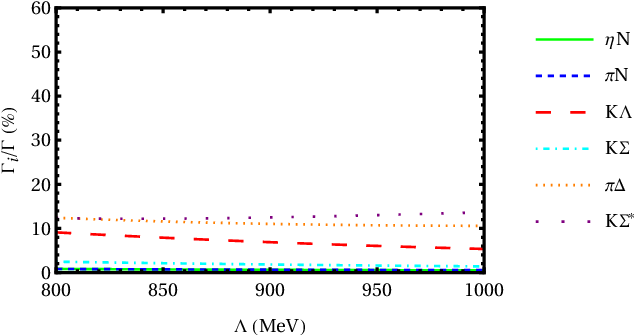}} {\hglue 0.32cm}
    \subfigure{
    \includegraphics[width=0.861\columnwidth]{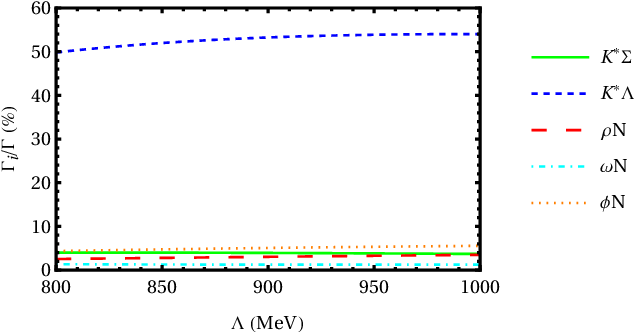}}
    \caption{Branching ratio for the $N(2270)3/2^-$ hadron decay as a function of cutoff. The notation is the same as Fig. \ref{fig:1h}.}
    \label{fig:3h}
\end{figure*}

\begin{figure*}[htb]
    \centering
    {\vglue 0.15cm}
    \subfigure{
    \includegraphics[width=0.855\columnwidth]{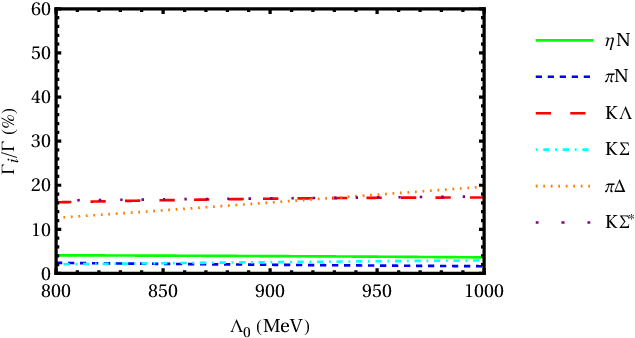}} {\hglue 0.4cm}
    \subfigure{
    \includegraphics[width=0.855\columnwidth]{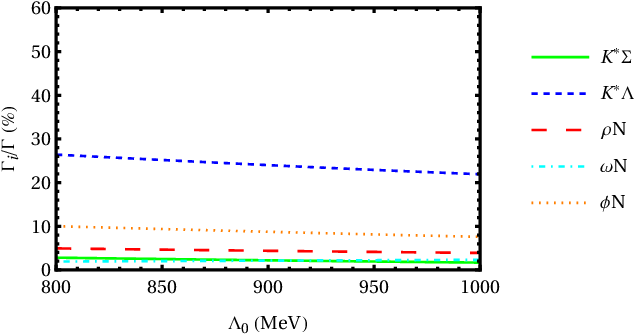}}\\
    \subfigure{
    \includegraphics[width=0.861\columnwidth]{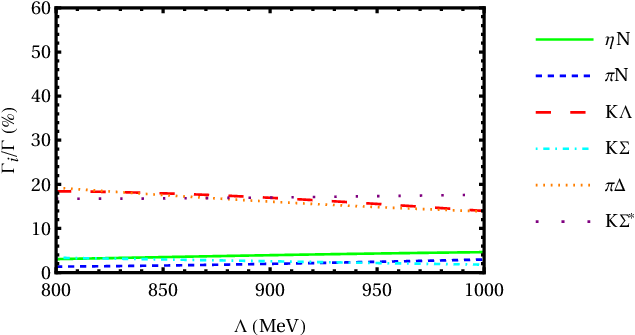}} {\hglue 0.32cm}
    \subfigure{
    \includegraphics[width=0.861\columnwidth]{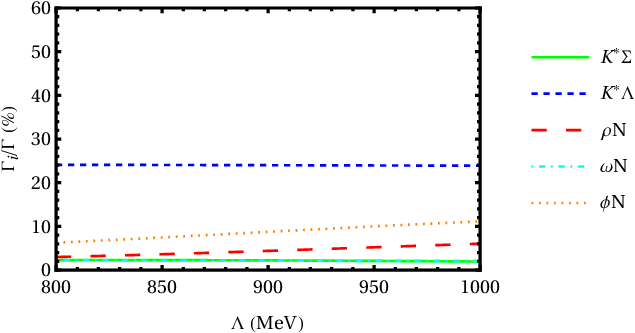}}
    \caption{Branching ratio for the $N(2270)5/2^-$ hadron decay as a function of cutoff. The notation is the same as Fig. \ref{fig:1h}.}
    \label{fig:5h}
\end{figure*}

As shown in Figs. \ref{fig:1h}, \ref{fig:3h} and \ref{fig:5h}, the graph in the left panel is the result of a fixed $\Lambda=900$ MeV as a function of $\Lambda_0$, while the graph in the right panel is the result of a fixed $\Lambda_0=900$ MeV as a function of $\Lambda$. At the same time, we draw the BRs of the decay channels of the final states containing pseudo-scalar mesons in the first row, where the green lines, blue dashed lines, red large-dashed lines, cyan dot-dashed lines, orange dotted lines, and purple large-dotted lines represent $\eta N$, $\pi N$, $K\Lambda$, $K\Sigma$, $\pi \Delta$ and $K\Sigma^\ast$ final state respectively, and all the cases of involving vector mesons in the second row, where the green lines, blue dashed lines, red large-dashed lines, cyan dot-dashed lines and orange dotted lines represent $K^\ast \Sigma$, $K^\ast \Lambda$, $\rho N$, $\omega N$ and $\phi N$ final state respectively.

When the cutoff parameter varies in a large range, which also corresponds to a large range of variation in the total width, besides the fact that $K\Lambda$ final state indicates to have a large dependence on both cutoffs $\Lambda_0$ and $\Lambda$ in $N(2270)1/2^-$ decays; and $K\Lambda$ and $\pi\Delta$ final states indicate to have a large dependence on both cutoffs $\Lambda_0$ and $\Lambda$ in $N(2270)5/2^-$ decays, the dependence of the rest of the decay channels on the varies of both cutoffs is not that obvious.

Taking $N(2270)1/2^-$ decays as an example, although $K\Lambda$ final state indicates to have a large shifting on both cutoffs $\Lambda_0$ and $\Lambda$, but what's important most is, it still show its absolute dominance in the decay patten of $N(2270)1/2^-$ decays, which means that this kind of variation does not influence the main conclusion. 
This discussion can also be applied to the case of $N(2270)5/2^-$ decays. To be more rigorous, for this situation, we give a range as the prediction in conclusion. 

For $N(2270)1/2^-$ hadronic decays, the main decay channels include $K\Lambda$, $K\Sigma$, $K \Sigma^\ast$, and $K^\ast \Lambda$. 
For the $K\Lambda$ final state, the BR is the largest, showing significant dependence on the cutoff values; the result varies in the range of $ 25\%\sim48\%$.
For $K\Sigma$ final state, the BR varies in the range of $10\% \sim 15\%$, while for the $K^\ast\Lambda$, $\rho N$, and $\phi N$ final states, their BRs are around $10\%$.

For $N(2270)3/2^-$ hadronic decays, the main decay channels are $K^\ast \Lambda$, $K \Sigma^\ast$, and $\pi\Delta$ final states. 
For $K^\ast \Lambda$ final state, the BR is the largest with $\sim 53\%$, while for $K \Sigma^\ast$ and $\pi\Delta$ final states, their BRs are around $12\%$.

For $N(2270)5/2^-$ hadronic decays, the main decay channels include $K^*\Lambda$, $K\Sigma^*$, $K\Lambda$ , $\pi\Delta$ and $\phi N$ final states. 
The largest BR $\sim 25\%$ comes out to be for the $K^\ast\Lambda$ final state, the next one is BR $\sim 17\%$ for $K\Sigma^*$. For $\pi\Delta$ and $K\Lambda$ final states, we obtian BR in the range of $14\%\sim20\%$, found to be quite sensitive to the cutoff value. The BR for $\phi N$ final state is around $10\%$.


\subsubsection{Radiative Decay}

\begin{figure}[tb]
    \centering
    {\vglue 0.15cm}
    \subfigure{
    \includegraphics[width=0.815\columnwidth]{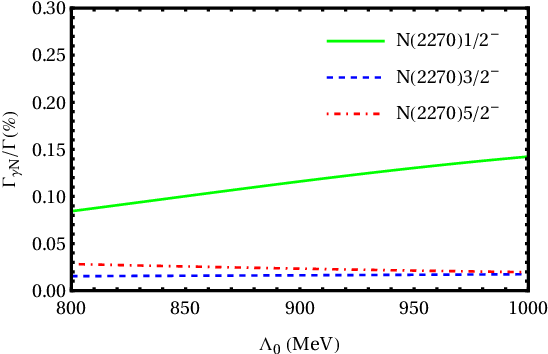}}
    \subfigure{
    \includegraphics[width=0.815\columnwidth]{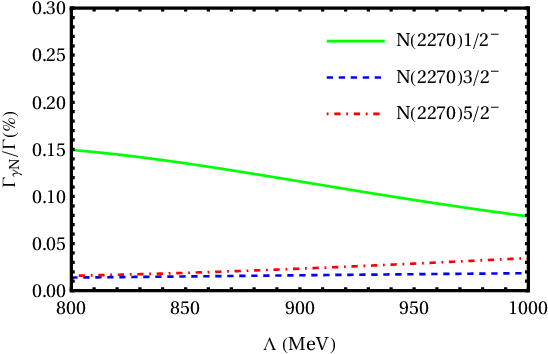}}
    \caption{Branching ratio of radiative decay as a function of cutoff, with $\Lambda=900$ MeV (upper figure) and $\Lambda_0=900$ MeV (lower figure). Where the green line, blue dashed line, and red dot-dashed line correspond to the case of initial state $N(2270)$ spin-parity quantum number $J^P$ as $1/2^-$, $3/2^-$, and $5/2^-$, respectively.}
    \label{fig:photon}
\end{figure}

The calculated BR of the radiative decay is shown in Fig. \ref{fig:photon}, where the upper graph is the result of fixing the $\Lambda$ as a function of $\Lambda_0$, and the lower graph is the result of fixing $\Lambda_0$ as a function of $\Lambda$. It shows that the BRs are of the order of $0.1\%$ or much smaller.

For the $J^P = 1/2^-$ case, the radiative decay BR indicates to have a large dependence on both cutoffs $\Lambda_0$ and $\Lambda$. To be more rigorous, we find that this radiative decay BR is $0.09\% \sim 0.15\%$.

For $J^P = 3/2^-$ or $5/2^-$ cases, we find that this radiative decay BR shows a smooth dependence on the cutoff, and we can confirm our calculated BR is around $0.02\%$ to be a reliable physical result.

\subsection{Discussion Independent of Cutoff}

For all three cases, the $K^\ast \Lambda$ final state is always included as the main decay channel.
This is interesting that from the picture of $N(2270)$ hadronic molecular decays, one of the components $\Sigma^*$ is most probable to hadronic decay to $\Lambda\pi$ systems~\cite{ParticleDataGroup:2024cfk}.
Such a significant feature is brought from the hadronic molecular picture.
Meanwhile, all three cases are also different from each other. 

Note that the other component $K^*$ is most probable to hadronic decay to the $K\pi$ system. In that case, $K\Sigma^\ast$ should also be raised as the main decay channel from the hadronic molecular picture. Actually, the $K\Sigma^\ast$ channel is suppressed according to the partial wave analysis, and the detailed discussion will be expanded below.

For simplicity, we abbreviate the final states with a pseudo-scalar meson and a baryon belonging to the baryon-octet as $PB$ final states. Also, the final states with a vector meson and a baryon belonging to the baryon-octet will be named as $VB$ final states, and the final states with a pseudo-scalar meson and a baryon belonging to the baryon-decuplet named $PD$ final states.

Comparing the decay pattern of $N(2270)3/2^-$ with $N(2270)1/2^-$, we can find that the suppression of some $PB$ final states including $K\Sigma$ and $K\Lambda$ in the main decay channel of the original $N(2270)1/2^-$, while the $K\Sigma^\ast$ and $\pi \Delta$ final state which is a $PD$ final state in $N(2270)3/2^-$ decay pattern become the new main decay channel, which is also very easy to understand, that is due to the difference in the quantum numbers of the initial states.

When the initial state spin parity $J^P = 3/2^-$, the $PB$ final states need d-wave couplings when they interact to $N(2270)3/2^-$ and thus get depressed, while the $PD$ final states are s-wave coupled to $N(2270)3/2^-$. In contrast, in the $J^P = 1/2^-$ case it is exactly the opposite, the $PD$ final states need d-wave couplings when they interact to $N(2270)1/2^-$ and the $PB$ final states are s-wave coupled to $N(2270)1/2^-$, that is why the $PD$ final states become one of the main decay channels in $N(2270)3/2^-$ decay pattern. Note that the $VB$ final states will not be suppressed in both $J^P = 1/2^-$ and $J^P = 3/2^-$ cases.

But, in the case of $J^P = 5/2^-$, no matter whether $PB$, $VB$, or $PD$ final states need at least a d-wave coupling, a new change in the overall decay pattern emerges.
At the same time, it is understandable why, with the same choice of cutoff parameters, the calculated total width of the $N(2270)5/2^-$ state is significantly smaller than that in the $N(2270)1/2^-$ and $N(2270)3/2^-$ cases.

\section{Summary and Conclusions} \label{sec:summary}

In this article, we systematically calculate the decay behavior of the hadronic molecular state $N(2270)$, which is composed of s-wave $K^\ast\Sigma^\ast$ with $J^P = 1/2^-$, $3/2^-$, and $5/2^-$, and discuss the dependence of the results on the cutoff parameters.

For all three cases, the $K^\ast \Lambda$ final state is always included as the main decay channel. Meanwhile, the $K\Sigma^*$ final state is also an outstanding one, although it is suppressed according to the partial wave analysis in the $J^P = 1/2^-$ case.
These features strongly indicate that the decay pattern of the hadronic molecular states is inextricably linked to the decay pattern of its constituent hadrons. 
However, the $K \Sigma$, $K \Lambda$, $K\Sigma^\ast$ and $\pi \Delta$ final states exhibit notable differences for different $J^P$.
The results we have obtained about the decay behavior will hopefully aid in the experimental discovery as well as the identification of the existence of $N(2270)$. 

Possible roles played by the $N(2270)3/2^{-}$ in the reactions $\gamma p \to K^* \Sigma$ \cite{Ben:2023uev} and $\gamma p \to p \phi$ \cite{Wu:2023ywu}, is compatible with the available cross-section data for the $\gamma p \to K^{*0} \Sigma^+$ channel \cite{CLAS:2007kab} and the process \cite{CLAS:2013qgi} $\gamma p \to K^{*+} \Sigma^0$.
In the $\gamma p \to p \phi$ process, a bump structure occurs \cite{Dey:2014tfa} around $W = 2.27$ GeV, which is explained \cite{Wu:2023ywu} by the $N(2270)3/2^-$.
In those investigations, only $J^P = 3/2^-$ was considered. In the present paper, we extend the domain from $J^P = 1/2^-$ to $J^P = 5/2^-$, and systematically study the decay patterns of the hidden-strange hadronic molecular state $N(2270)$ which is assumed to be an s-wave $K^*\Sigma^+$ shallow bound state, the strange partner of the $D^*\Sigma^*$ bound states.
Our findings could be tested by measurements at JLab \cite{jlab} and J-PARC \cite{j-parc} via various channels; namely, $\gamma p$, $\pi p \to K^*\Lambda$, $K\Sigma^*$, $K\Lambda$, $K\Sigma$ and $\pi\Delta$, as well as at high-energy under-construction facility, HIAF \cite{hiaf}, via the reactions such as $pp \to pp\phi$, $p K^*\Lambda$, $p K^*\Sigma$, $pK\Sigma^*$, to which our formalism could be applied.

Furthermore, we show that the $K^*\Lambda$ channel has a significant impact on the formation of the $K^*\Sigma^*$ molecular state.
In the future, we will investigate the full coupled-channel effects and conduct a more detailed study of the decay process of the $N(2270)$.


\begin{acknowledgments}
Thanks for the useful discussion with Professor Bing-Song Zou, Feng-Kun Guo, Jia-Jun Wu, and Fei Huang. This work is partially supported by the National Natural Science Foundation of China under Grants No. 12175240 and No. 11635009, by the Fundamental Research Funds for the Central Universities, and by the NSFC and the Deutsche Forschungsgemeinschaft (DFG, German Research Foundation) through the funds provided to the Sino-German Collaborative Research Center TRR110 “Symmetries and the Emergence of Structure in QCD” (NSFC Grant No. 12070131001, DFG Project-ID 196253076 - TRR 110), by the NSFC Grant No.11835015, No.12047503, No. 12175239, No. 12221005 and by the Chinese Academy of Sciences (CAS) under Grant No.XDB34030000, and by the Chinese Academy of Sciences under Grant No. YSBR-101.
\end{acknowledgments}


\begin{thebibliography}{99}
%
\bibitem{Aaij:2015tga}
R. Aaij {\it et al.} (LHCb Collaboration), Phys. Rev. Lett. {\bf 115}, 072001 (2015).
%
\bibitem{Aaij:2019}
R. Aaij {\it et al.} (LHCb Collaboration), Phys. Rev. Lett. {\bf 122}, 222001 (2019).
%
\bibitem{Wu:2010jy}
J.~J.~Wu, R.~Molina, E.~Oset and B.~S.~Zou,
Phys. Rev. Lett. \textbf{105}, 232001 (2010)
\bibitem{Wu:2010vk}
J.~J.~Wu, R.~Molina, E.~Oset and B.~S.~Zou,
Phys. Rev. C \textbf{84}, 015202 (2011)
\bibitem{Wang:2011rga}
W.~L.~Wang, F.~Huang, Z.~Y.~Zhang and B.~S.~Zou,
Phys. Rev. C \textbf{84}, 015203 (2011)
\bibitem{Yang:2011wz}
Z.~C.~Yang, Z.~F.~Sun, J.~He, X.~Liu and S.~L.~Zhu,
Chin. Phys. C \textbf{36}, 6-13 (2012)
\bibitem{Wu:2012md}
J.~J.~Wu, T.~S.~H.~Lee and B.~S.~Zou,
Phys. Rev. C \textbf{85}, 044002 (2012)
\bibitem{Yuan:2012wz}
S.~G.~Yuan, K.~W.~Wei, J.~He, H.~S.~Xu and B.~S.~Zou,
Eur. Phys. J. A \textbf{48}, 61 (2012)
\bibitem{Xiao:2013yca}
C.~W.~Xiao, J.~Nieves and E.~Oset,
Phys. Rev. D \textbf{88}, 056012 (2013)
\bibitem{Uchino:2015uha}
T.~Uchino, W.~H.~Liang and E.~Oset,
Eur. Phys. J. A \textbf{52}, no.3, 43 (2016)
\bibitem{Karliner:2015ina}
M.~Karliner and J.~L.~Rosner,
Phys. Rev. Lett. \textbf{115}, no.12, 122001 (2015)
%
\bibitem{Chen:2016qju}
H. X. Chen, W. Chen, X. Liu, and S. L. Zhu, Phys. Rept. {\bf 639}, 1 (2016).
\bibitem{Ali:2016dkf}
A.~Ali, I.~Ahmed, M.~J.~Aslam and A.~Rehman,
Phys. Rev. D \textbf{94}, no.5, 054001 (2016)
\bibitem{Ali:2019npk}
A.~Ali and A.~Y.~Parkhomenko,
Phys. Lett. B \textbf{793}, 365-371 (2019)
\bibitem{Maiani:2015vwa}
L.~Maiani, A.~D.~Polosa and V.~Riquer,
Phys. Lett. B \textbf{749}, 289-291 (2015)
\bibitem{Li:2015gta}
G.~N.~Li, X.~G.~He and M.~He,
JHEP \textbf{12}, 128 (2015)
\bibitem{Mironov:2015ica}
A.~Mironov and A.~Morozov,
JETP Lett. \textbf{102}, no.5, 271-273 (2015)
\bibitem{Anisovich:2015cia}
V.~V.~Anisovich, M.~A.~Matveev, J.~Nyiri, A.~V.~Sarantsev and A.~N.~Semenova,

\bibitem{Zhu:2015bba}
R.~Zhu and C.~F.~Qiao,
Phys. Lett. B \textbf{756}, 259-264 (2016)
\bibitem{Ghosh:2015xqp}
R.~Ghosh, A.~Bhattacharya and B.~Chakrabarti,
Phys. Part. Nucl. Lett. \textbf{14}, no.4, 550-552 (2017)
\bibitem{Wang:2015epa}
Z.~G.~Wang,
Eur. Phys. J. C \textbf{76}, no.2, 70 (2016)
\bibitem{Hiyama:2018ukv}
E.~Hiyama, A.~Hosaka, M.~Oka and J.~M.~Richard,
Phys. Rev. C \textbf{98}, no.4, 045208 (2018)
%
\bibitem{Guo:2017jvc}
F. K. Guo, C. Hanhart, U.-G. Mei{\ss}ner, Q. Wang, Q. Zhao, and B. S. Zou, Rev. Mod. Phys. {\bf 90}, 015004 (2018).
\bibitem{He:2019rva}
J.~He and D.~Y.~Chen,
Eur. Phys. J. C \textbf{79}, no.11, 887 (2019)
\bibitem{Chen:2019asm}
R.~Chen, Z.~F.~Sun, X.~Liu and S.~L.~Zhu,
Phys. Rev. D \textbf{100}, no.1, 011502 (2019)
\bibitem{Liu:2019zvb}
M.~Z.~Liu, T.~W.~Wu, M.~S\'anchez S\'anchez, M.~P.~Valderrama, L.~S.~Geng and J.~J.~Xie,
Phys. Rev. D \textbf{103}, no.5, 054004 (2021)
\bibitem{Du:2021fmf}
M.~L.~Du, V.~Baru, F.~K.~Guo, C.~Hanhart, U.~G.~Mei\ss{}ner, J.~A.~Oller and Q.~Wang,
JHEP \textbf{08}, 157 (2021)
\bibitem{Yalikun:2021bfm}
N.~Yalikun, Y.~H.~Lin, F.~K.~Guo, Y.~Kamiya and B.~S.~Zou,
Phys. Rev. D \textbf{104}, no.9, 094039 (2021)
%
\bibitem{Wu:2024bvl}
J.~Z.~Wu, J.~Y.~Pang and J.~J.~Wu,
Chin. Phys. Lett. \textbf{41}, no.9, 091201 (2024)
\bibitem{He:2017aps}
J. He, Phys. Rev. D \textbf{95}, 074031 (2017).
%
\bibitem{Lin:2018kcc}
Y. H. Lin, C. W. Shen, and B. S. Zou, Nucl. Phys. A \textbf{980}, 21 (2018).
%

%
\bibitem{Ben:2023uev}
D.~Ben, A.~C.~Wang, F.~Huang and B.~S.~Zou, Phys. Rev. C \textbf{108}, no.6, 065201 (2023).
%
\bibitem{Wu:2023ywu}
S.~M.~Wu, F.~Wang and B.~S.~Zou, Phys. Rev. C \textbf{108}, no.4, 045201 (2023).
%
\bibitem{Zou:2002yy}
B.~S.~Zou and F.~Hussain,
Phys. Rev. C \textbf{67}, 015204 (2003)
\bibitem{Ronchen:2012eg}
D.~Ronchen, M.~Doring, F.~Huang, H.~Haberzettl, J.~Haidenbauer, C.~Hanhart, S.~Krewald, U.~G.~Meissner and K.~Nakayama,
Eur. Phys. J. A \textbf{49}, 44 (2013)
doi:10.1140/epja/i2013-13044-5
%
\bibitem{Oset:2010tof}
E.~Oset and A.~Ramos, Eur. Phys. J. A \textbf{44}, 445-454 (2010)
%
\bibitem{Matsuyama:2006rp}
A.~Matsuyama, T.~Sato and T.~S.~H.~Lee,
Phys. Rept. \textbf{439}, 193-253 (2007)
%
\bibitem{Lin:2019qiv}
Y.~H.~Lin and B.~S.~Zou,
Phys. Rev. D \textbf{100}, no.5, 056005 (2019)
doi:10.1103/PhysRevD.100.056005

\bibitem{Meng:2017fwb}
L.~Meng, N.~Li and S.~L.~Zhu,
Phys. Rev. D \textbf{95}, no.11, 114019 (2017)

\bibitem{Nakayama:2006ps}
K.~Nakayama, Y.~Oh, J.~Haidenbauer and T.~S.~H.~Lee,
Phys. Lett. B \textbf{648}, 351-356 (2007)

\bibitem{Janssen:1996kx}
G.~Janssen, K.~Holinde and J.~Speth,
Phys. Rev. C \textbf{54}, 2218-2234 (1996)
doi:10.1103/PhysRevC.54.2218

\bibitem{Janssen:1994wn}
G.~Janssen, B.~C.~Pearce, K.~Holinde and J.~Speth,
Phys. Rev. D \textbf{52}, 2690-2700 (1995)
doi:10.1103/PhysRevD.52.2690

\bibitem{Schutz:1998jx}
C.~Schutz, J.~Haidenbauer, J.~Speth and J.~W.~Durso,
Phys. Rev. C \textbf{57}, 1464-1477 (1998)
doi:10.1103/PhysRevC.57.1464

\bibitem{Lu:2024ajt}
Y.~Lu, H.~J.~Jing and J.~J.~Wu,
Symmetry \textbf{16}, no.8, 1061 (2024)
doi:10.3390/sym16081061

\bibitem{deSwart:1963pdg}
J.~J.~de Swart,
Rev. Mod. Phys. \textbf{35}, 916-939 (1963)
[erratum: Rev. Mod. Phys. \textbf{37}, no.2, 326-326 (1965)]
doi:10.1103/RevModPhys.35.916

\bibitem{ParticleDataGroup:2024cfk}
S.~Navas \textit{et al.} [Particle Data Group],
Phys. Rev. D \textbf{110}, no.3, 030001 (2024)
doi:10.1103/PhysRevD.110.030001

\bibitem{Wu:2019adv}
J.~J.~Wu, T.~S.~H.~Lee and B.~S.~Zou,
Phys. Rev. C \textbf{100}, no.3, 035206 (2019)

\bibitem{CLAS:2007kab}
I.~Hleiqawi \textit{et al.} [CLAS],
Phys. Rev. C \textbf{75}, 042201 (2007)
[erratum: Phys. Rev. C \textbf{76}, 039905 (2007)]
doi:10.1103/PhysRevC.76.039905
[arXiv:nucl-ex/0701036 [nucl-ex]].

\bibitem{CLAS:2013qgi}
W.~Tang \textit{et al.} [CLAS],
Phys. Rev. C \textbf{87}, no.6, 065204 (2013)
doi:10.1103/PhysRevC.87.065204
[arXiv:1303.2615 [nucl-ex]].

\bibitem{Dey:2014tfa}
B.~Dey \textit{et al.} [CLAS],
Phys. Rev. C \textbf{89}, no.5, 055208 (2014)
doi:10.1103/PhysRevC.89.055208
[arXiv:1403.2110 [nucl-ex]].

\bibitem{jlab}
\url{https://www.jlab.org/}

\bibitem{j-parc}
\url{https://j-parc.jp/c/en/}

\bibitem{hiaf}
\url{https://hiaf.impcas.ac.cn/hiaf_en/public/}

\end{thebibliography}
\end{document}